# The Roles of Supervised Machine Learning in Systems Neuroscience

Joshua I. Glaser[1]*, Ari S. Benjamin[1]*, Roozbeh Farhoodi[1]*, Konrad P. Kording[1,2]

1. Department of Bioengineering, University of Pennsylvania
2. Department of Neuroscience, University of Pennsylvania, Member, Canadian Institute For Advanced Research

*These authors contributed equally to this work.

Contacts: jglaser8, aarrii, roozbeh, kording @seas.upenn.edu

## Abstract:

Over the last several years, the use of machine learning (ML) in neuroscience has been rapidly increasing. Here, we review ML's contributions, both realized and potential, across several areas of systems neuroscience. We describe four primary roles of ML within neuroscience: 1) creating solutions to engineering problems, 2) identifying predictive variables, 3) setting benchmarks for simple models of the brain, and 4) serving itself as a model for the brain. The breadth and ease of its applicability suggests that machine learning should be in the toolbox of most systems neuroscientists.

## Introduction:

There is a lot of enthusiasm about machine learning (ML). After all, it has allowed computers to surpass human-level performance at image classification (He et al. 2015), to beat humans in complex games such as "Go" (Silver et al. 2016), and to provide high-quality speech to text (Hannun et al. 2014) in popular mobile phones. Progress in ML is also getting attention in the scientific community. Writing in the July 2017 issue of Science focusing on "AI Transforms Science", editor Tim Appenzeller writes, "For scientists, prospects are mostly bright: AI promises to supercharge the process of discovery" (Appenzeller 2017).

The field of systems neuroscience is no exception. In the last few years there have been many opinion pieces about the importance of ML in neuroscience (Vu et al. 2018; Barak 2017; Paninski and Cunningham 2017; Vogt 2018; Hinton 2011). Moreover, when we analyze the number of journal articles about ML in neuroscience, we find that its use has been continuously growing over the last 20 years (Fig. 1). Machine learning has been used in many different ways within this literature. In this review, we will catalog the conceptual applications of ML in systems neuroscience.

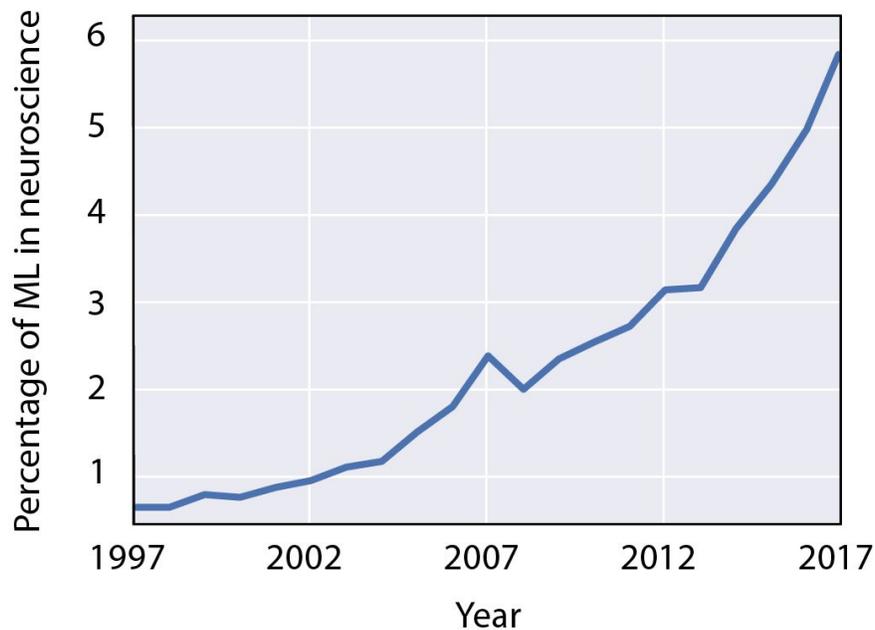

**Figure 1: Growth of Machine Learning in Neuroscience.**
Here we plot the proportion of neuroscience papers that have used ML over the last two decades. That is, we calculate the number of papers involving both neuroscience and machine learning, normalized by the total number of neuroscience papers. Neuroscience papers were identified using a search for "neuroscience" on Semantic Scholar. Papers involving neuroscience and machine learning were identified with a search for "machine learning" and "neuroscience" on Semantic Scholar.

On the highest level, ML is typically divided into the subtypes of supervised, unsupervised, and reinforcement learning. Supervised learning builds a model that predicts outputs from input data. Unsupervised learning is concerned with finding structure in data, e.g. clustering, dimensionality reduction, and compression. Reinforcement learning allows a system to learn the best actions based on the reward that occurs at an end of a sequence of actions. This review focuses on supervised learning.

Why is creating progressively more accurate regression or classification methods (see Box 1) worthy of a title like 'The AI Revolution' (Appenzeller 2017)? It is because countless questions can be framed in this manner. When classifying images, an input picture can be used to predict the object in the picture. When playing a game, the setup of the board (input) can be used to predict an optimal move (output). When texting on our smartphones, our current text is used to create suggestions of the next word. Similarly, science has many instances where we desire to make predictions from measured data.

In this review, we categorize the ways in which supervised ML promises to assist, or has already been applied to, problems in systems neuroscience. We believe that applications of supervised ML in this field can be divided in roughly four categories (Fig. 2). 1) *Solving engineering problems*. Machine learning can improve the predictive performance of methods used by neuroscientists, such as medical diagnostics, brain-computer interfaces, and research tools. 2) *Identifying predictive variables*. Machine learning can more accurately determine whether variables (e.g., those related to the brain and outside world) predict each other. 3) *Benchmarking simple models*. We can compare the performance of simple interpretable models to highly accurate ML models in order to help determine the quality of the simple models. 4) *Serving as a model for the brain.* We can argue whether the brain solves problems in a similar way to ML systems, e.g. deep neural networks. The logic behind each of these applications is rather distinct.

For the bulk of this review, we will go into further detail about these four roles of ML in neuroscience. We will provide many examples, both realized and potential, of ML across several areas of systems neuroscience. More specifically, we will discuss the four roles of ML in relation to *neural function*, including neural activity and how it relates to behavior; and *neural structure*, i.e., neuroanatomy. We also discuss ML in practice (Box 2), as that is crucial for useful applications.

---

**Box 1: A very brief glossary of supervised ML terms**

**Regression:** Prediction of continuous-valued output (e.g., the amount of amyloid plaques in the brain)

**Classification:** Prediction of categorical output (e.g., whether a subject has Alzheimer's disease)

**K-nearest neighbors:** This set of classification and regression methods predict an output by comparing to the *k* closest data points in input space.

**Generalized linear model (GLM):** A regression or classification method in which features are first linearly combined, and then fed through an output function (e.g., an exponential function).

**Logistic regression:** One type of GLM, that uses a logistic (sigmoid) function as an output nonlinearity, so that outputs between 0 and 1 are predicted. It is used for classification.

**Support vector machine (SVM):** A classification method that (often nonlinearly) projects the inputs into a high-dimensional space and then creates boundaries to divide data in this high-dimensional space.

**Artificial neural networks (deep learning):** These methods map input to output through a network of nodes, each of which affects the others with a learnable weight. *Feedforward neural networks* repeatedly transform the data (across multiple layers of the neural net) using linear combinations followed by output nonlinearities. *Recurrent neural networks (RNNs)* allow nodes to excite themselves, or are generally cyclic, and are generally used for sequential data (e.g., time series, speech data). *Convolutional neural networks (CNNs)* are most often used for images, and learn filters that are applied in the same way to all parts of the input. This allows the network to learn features of the image regardless of their precise location within the image.

**Tree-based methods:** Classification and regression trees are decision trees that, like a flow chart, learn to sequentially split the input space via decision boundaries on each variable. The final divisions of input space ("leaves") are assigned output values. Common tree-based methods use "ensembles" (see below) of trees. Random forests average across many trees. XGBoost fits new trees to the residuals of previously fit trees.

**Ensembles:** Ensemble methods combine the predictions from many different models. They can average the results across models trained on resampled data (bagging), sequentially fit models based on the errors of previous models (boosting), or feed the initial model predictions into an additional ML model (stacking).

**Bootstrapping:** A common method for getting confidence intervals of model parameters or estimators. The original dataset is resampled, with replacement, to create new "bootstrapped" datasets, which are used to generate confidence intervals.

## 1 - Solving Engineering Problems

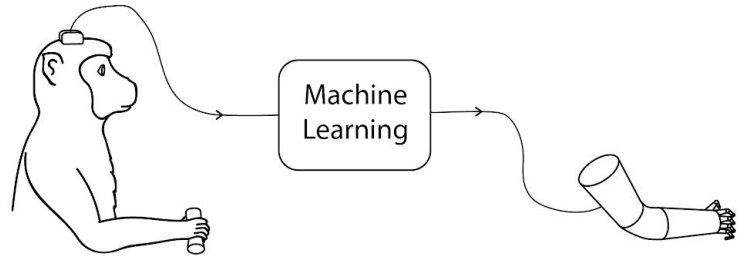

## 2 - Identifying Predictive Variables

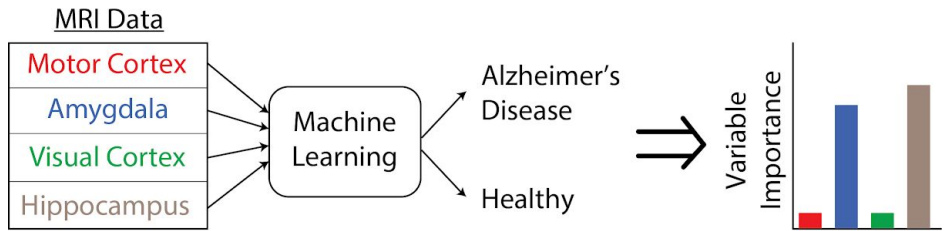

## 3 - Benchmarking Simple Models

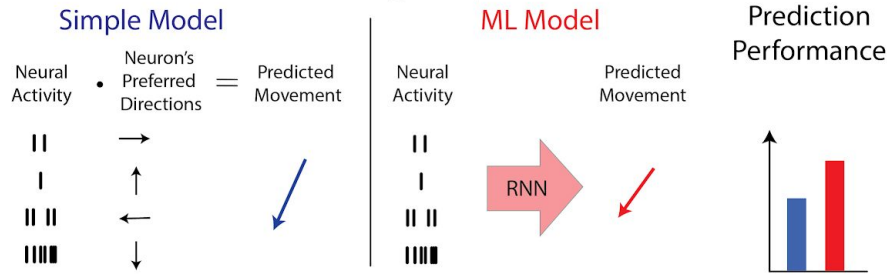

## 4 - Serving as a Model for the Brain

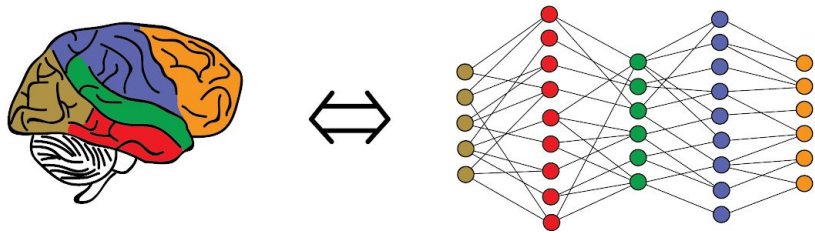

**Figure 2: Examples of the four roles of supervised machine learning in neuroscience.**
1 - *ML can solve engineering problems*. For example, it can help researchers control a prosthetic limb using brain activity. 2 - *ML can identify predictive variables*. For example, by using MRI data, we can identify which brain regions are most predictive for diagnosing Alzheimer's disease (Lebedev et al. 2014). 3 - *ML can benchmark simple models*. For example, we can compare the predictive performance of the simple "population vector" model of how neural activity relates to movement (Georgopoulos, Schwartz, and Kettner 1986) to a ML benchmark (e.g. an RNN). 4 - *ML can serve as a model of the brain*. For example, researchers have studied how neurons in the visual pathway correspond to units in an artificial network that is trained to classify images (Yamins and DiCarlo 2016).

# Role 1: Solving engineering problems

A surprisingly wide array of engineering problems can be cast as prediction problems. Their common thread is that one would like to estimate some quantity of interest (**Y**) and can take measurements (**X**) that relate to that quantity. The relationship between **X** and **Y**, however, is unknown and might be complicated. We call these 'engineering problems' when the final quantity, **Y**, is all that is desired. In these problems, one does not need detailed understanding of the relationship - the aim is simply to estimate **Y** as accurately as possible. For example, email providers wish to filter spam from incoming messages, and only care about how accurately the emails are sorted.

Traditionally, one would attempt to carefully understand the relationship between **X** and **Y**, and synthesize this into a model. Modern machine learning (ML) is changing this paradigm. Instead of detailed expert knowledge of a process, a practitioner simply needs a large database of measurements along with the associated quantity of interest for each measurement. Machine learning algorithms can then automatically model their relationship. Once trained, an ML algorithm can make predictions for new measurements.

This 'engineering' framework is the traditional application of ML and is common in industry. In neuroscience, there are many problems that resemble this general problem format.

## Neural Activity / Function

Many medical applications depend on successfully extracting information about intention, sensation, or disease from measurements of neural activity. This is a difficult problem, since the meaning of neural activity, the 'neural code', is most often not known. Machine learning is now a ubiquitous tool for this task in situations in which it is possible to obtain large datasets of neural activity along with the behaviors or diseases of interest. One such application is brain-computer interfaces (BCIs), which seek to use neural signals to control prosthetic limbs, computer cursors, or other external objects. Several groups have used modern ML techniques, such as recurrent neural networks, to improve BCIs using recordings of spikes (Sussillo et al. 2016, 2012), ECoG (Elango et al. 2017), or EEG (Bashivan et al. 2015). Machine learning can also be used to predict future neural activity from past neural activity. This application is relevant for epilepsy, since imminent seizures can be predicted using deep learning (Talathi 2017; Nigam and Graupe 2004) and ensemble methods (Brinkmann et al. 2016). In another line of applications, researchers have used ML to diagnose neurological conditions from activity. Several reviews on this specific application have recently been published (see (Arbabshirani et al. 2017) for classification using neuroimaging data, (Rathore et al. 2017) for classification of Alzheimer's disease, and (Vieira, Pinaya, and Mechelli 2017) for a focus on deep learning approaches). Due to the large datasets available for neural recordings, ML has improved the standards of accuracy on these medical applications despite the complexity of the input signals.

An ML approach also promises to assist with the inverse of the above problem: predicting neural activity from variables in the outside world. Solving this problem is important if we want to use neural stimulation to induce accurate sensation. A prosthetic eye, for example, could be built by stimulating retinal ganglion cells in the correct manner according to the output of a camera (Nirenberg and Pandarinath 2012). The most accurate model of ganglion cell activity is currently a deep learning model trained to predict activity from naturalistic scenes (McIntosh et al. 2016). Similarly, prosthetic limbs could provide tactile and proprioceptive sensations if somatosensory neurons were correctly stimulated (Armenta Salas et al. 2018). Machine learning models may help to enable these neural prosthetics to induce sensations.

A very similar approach can be used to quantify behavior (D. J. Anderson and Perona 2014), such as movement, sleeping, and socializing. For example, we may want to quantify movement of the whole body using cheap video recordings. Recent progress in the field has made video quantification far more precise. Researchers have used deep learning to estimate human poses from video (Insafutdinov et al. 2016). Related approaches have recently gotten easier to use and less data intensive, and have been extended to animal tracking (Mathis et al. 2018; T. Pereira et al. 2018). Along with estimating poses, we can also directly estimate types of behavior (e.g., walking vs. stopping vs. jumping) from video (Kabra et al. 2013). Behavior can also be estimated based on other modalities such as audio recordings (Arthur et al. 2013). This progress in ML-based behavior tracking is timely, given recent calls to understand neural control of more naturalistic behavior in more natural environments (Krakauer et al. 2017).

The engineering approach of applying ML is also helping solve the problem of obtaining accurate estimates of neural activity from raw measurements. Many imaging methods, such as EEG, MEG, and fMRI, require the solving of an 'inverse problem' – obtaining the source from the measurements. For example, researchers estimate the origin of EEG signals within the brain based on electrode recordings from the scalp. Recently, it has been observed that deep learning can improve the estimates for imaging (McCann, Jin, and Unser 2017). Neural networks have improved image denoising (Burger, Schuler, and Harmeling 2012; Xie, Xu, and Chen 2012) and deconvolution (L. Xu et al. 2014), can provide super-resolution images (Dong et al. 2016), and can even replace the entire image processing pipeline (Golkov et al. 2016; Vito et al. 2005). Outside of imaging, the deconvolution of time series data is another common application. For example, once a researcher has obtained traces of cellular calcium concentration, there is still the difficult 'inverse problem' of inferring the timing of the underlying spiking. Competition-style ML on labeled datasets (Theis et al. 2016) provides good solutions (Berens et al. 2017). In each of these applications, a difficult engineering problem was replaced by building a large labeled dataset and using ML to learn the desired relationship.

**Neuroanatomy / Structure**

Just as neural activity can be indicative of disease, so can neuroanatomy. As such, it is often possible to take anatomical measurements and use machine learning to diagnose disease. For example, researchers can distinguish between Alzheimer's disease and healthy brains of older adults using MRI scans (Sarraf, Tofighi, and Others 2016). More generally, neuroanatomical measurements such as structural MRI and diffusion tensor imaging (DTI) can distinguish healthy from unhealthy patients across many conditions including schizophrenia, depression, autism and ADHD (Arbabshirani et al. 2017; Rathore et al. 2017; Vieira, Pinaya, and Mechelli 2017). Sometimes, ML enables surprising opportunities. For example, using deep convolutional neural networks, we can surprisingly predict cardiovascular risk factors from retinal fundus photographs (Poplin et al. 2018). The future will undoubtedly see ongoing efforts to automatically detect disease from biological data.

Since the majority of research in neuroanatomy is based on imaging techniques, recent advances in computer vision using ML are becoming important tools for neuroanatomy. Thus, segmenting and labeling parts of an image, which usually requires manual annotation, is an especially important task. However, as imaging techniques improve and the volume of data increases, it will become infeasible to rely on manual annotation. To solve this problem, many ML techniques have been developed to automatically segment or label new images based on a dataset of previously labeled images. The vast majority of techniques in this developing literature are based, at least in part, on convolutional neural networks (Litjens et al. 2017; Ronneberger, Fischer, and Brox 2015). This approach has been employed to label medical images, such as in identifying white matter tracts from MRI scans (Ghafoorian et al. 2017). They have also been used to understand the connections and morphologies of neurons from electron microscopy (Helmstaedter et al. 2013; Funke et al. 2017; Turaga et al. 2010). As imaging data improve in resolution and volume, ML is becoming a crucial and even necessary tool for reconstructing and mapping neuroanatomy.

**Caveats**

While ML methods have been able to provide many engineering solutions, ML is not magic (Wolpert and Macready 1997). Several conditions must be met for an ML method to solve a problem successfully.

A first consideration is that the selected method must match the structure of the data. For example, convolutional neural networks make the assumption that images have common local features (like edges), which allows them to be more successful than standard feedforward neural networks. In practice, this means that a user of ML must take care to select their method, or alternatively to preprocess their data, such that the assumptions match how the inputs relate to outputs. This requires good knowledge of the specifics of the data and also of ML methods. A workaround to this is to apply automated ML methods, which iterate intelligently through many possible model configurations and select the best-performing option (Guyon et al. 2015; Elsken, Metzen, and Hutter 2018; Feurer et al. 2015; Kotthoff et al. 2017). This approach is slow but often works comparatively well. The general rule, however, is that good ML engineering requires a good awareness of the data.

Another potential failure mode of ML is "overfitting" to the training data. Ideally, an ML method should accurately predict data it is not trained on. If a method learns to make accurate predictions on the training data but cannot generalize to new data, it is said to be overfit. To guard against the worry of overfitting, ML practitioners always report a model's performance on a test set that the model has not been trained on. The capacity of a method to overfit to data can be lowered with regularization methods, which penalize the complexity of a model. Still, overfitting is especially worrisome for small datasets and complex models. While different methods have different sensitivities to the number of data points, all methods become less vulnerable to overfitting when datasets are large. Sometimes simpler methods may be better choices on small datasets, even if a more complex method could better express the underlying input/output relationship (if there was sufficient data). The risk of overfitting means that all ML practitioners must be aware of regularization techniques, their dataset size, and the importance of reporting accuracy on a test set not used for training.

Yet another practical drawback of ML is that it can be slow. For large datasets and complex models, the time it takes to train the model can be prohibitive without proper hardware, like GPUs for deep learning. Once a model is trained, however, it is much faster to make predictions. Still, for applications that require real-time predictions, even the prediction step may be too slow for some ML methods. For example, predictions for brain machine interfaces often need to be made in the timescale of tens of milliseconds, which can be a challenge for models requiring many computations. This tradeoff between complexity and run-time is an important aspect in choosing a model for many engineers.

## Role 2: Identifying predictive variables

Neuroscientists often ask questions of the form, "which variables are related to something of interest?" For example, which brain regions can predict each other? Which brain regions contain information related to a subject's decision? Which cell types are affected by a certain disease? Machine learning (ML) can help to more accurately identify how informative one set of variables is about another. This is particularly instructive when there is a complex nonlinear relationship between the variables, which is often the case in neural systems. Answering these types of questions allows researchers to better understand the relationship between parts of the brain, stimuli, behavior, and more.

The general strategy resembles that of the engineering applications (Role 1). However, instead of only searching for maximal predictive accuracy, one examines which input variables lead to that accuracy. There are many methods to establish so-called 'feature importance' (also known as 'feature selection'). Two of the simplest are the leave-one-out strategy, in which each variable is removed and one observes the decrease in accuracy, and the 'best first' strategy, in which the algorithm is run on each variable alone. Leave-one-out reflects the information in that variable but not the others, while best-first reflects the total (learnable) task information in each variable. Both are related but different definitions of what it means for a feature to be important. The development of feature importance measures is an active field in ML and statistics (Tang,

Alelyani, and Liu 2014). These methods allow us to get insights into which variables are important for a given problem (with the specific meaning of 'importance' depending on the measure used).

A more traditional approach for this type of question would be to fit simple models to data, like linear regression, and to examine the coefficients of the fit. This approach is ubiquitous in science. Its basic drawback, however, is that one needs to assume a model, which may be inaccurate. For example, if the model is assumed to be y = mx+b, but the true relationship is y = cos x, then the value of m (the "interaction between x and y") will be 0 despite there being a strong relationship between x and y. The ML approach, on the other hand, seeks to maximize predictive accuracy and in doing so does not need to assume a simple functional form. This has the advantage that we can evaluate a variable's importance even when the relationship between inputs and outputs is unknown and possibly nonlinear. Plus, by bootstrapping, we can even find the confidence interval for their importance values. Machine learning combined with feature selection approaches can be universally applied to problems regardless of whether we know the underlying relationship.

Determining the important features can also help us to construct simpler models. Rather than using many inputs for a model, we can only use the important features as inputs. For example, determining which morphological features of neurons are most predictive of cell type can lead us to build more accurate generative models of morphologies (that are based on the most predictive features). Accurately determining the importance of features within ML algorithms is thus also beneficial for creating simpler models.

This approach is designed to examine the same variables that serve as raw inputs into the ML model. Often the "features" we seek are different than the raw inputs. In vision, for example, the raw input variables may be as simple as single pixels, and the approach of Role 2 would then simply isolate the most relevant pixels. However, as we review below, many problems in neuroscience follow the format where the input variables are the variables of interest.

**Neural Activity / Function**

Neuroscience has a long history of building encoding models, which aim to predict neural activity (e.g., spikes in an individual neuron, or BOLD signal in an fMRI voxel) based on variables in the outside world. This is a common approach to identifying the "role" of a brain area. The building of encoding models is a regression problem (from external variables to activity), and its purpose is more akin to feature importance than purely predictive power. This problem is an open invitation to use ML methods in combination with methods of determining feature importance.

Machine learning would not be necessary for encoding models if simpler methods were as accurate at describing neural activity. However, this is usually not the case. For example, we recently showed that XGBoost and ensemble methods led to significant performance improvements on datasets from motor cortex, somatosensory cortex, and hippocampus (Benjamin et al. 2018). These improvements were relative to Generalized Linear Models, which are ubiquitous in computational neuroscience. Others have also shown predictive improvements in other areas and modalities using methods such as XGBoost (Viejo, Cortier, and Peyrache 2018) and deep learning (McIntosh et al. 2016; Klindt et al. 2017; Agrawal et al. 2014). These instances serve as warnings that although simple models may appear interpretable, they may be missing important aspects of how external variables relate to neural function.

Having improved encoding performance can more generally allow researchers to understand which covariates are predictive of neural activity. This generalizes the already common approach of adding additional variables to simple models and observing the increase in performance (e.g., (Stringer et al. 2018)). For example, research on building encoding models of head-direction neurons using XGBoost (Viejo, Cortier, and Peyrache 2018) looked at the relative contribution of the different covariates (such as the direction of the head) within the encoding model. This allowed determining how the covariates mattered, without assuming the form of the relationship.

The reverse problem, "what information can be read-out from activity from this brain area" can also answer questions about information content and the role of specific brain areas or cell types. For example, researchers used decoding methods to compare how predictive neural populations in parietal and prefrontal cortices are about task-relevant variables, over the course of a decision-making task (Sarma et al. 2016). As another example, we have compared decoding results from motor cortex from different task conditions to determine how uncertainty in the brain relates to different behavioral uncertainty (Dekleva et al. 2016; Glaser et al. 2018). The choice of decoding method has a large impact on performance. We have recently done a thorough test of different ML methods on datasets from motor cortex, somatosensory cortex, and hippocampus, and have shown that modern ML methods, such as neural networks and ensemble methods lead to increased decoding accuracy (Glaser et al. 2017). More accurate decoding can increase our understanding of how much information is contained in a neural population about another variable, such as a decision, movement, or location.

Neuroscience researchers often want to determine which variables matter for behavior, so that they can relate these variables to neural activity. We can apply ML to find what variables are predictive of behavior, without assuming the form of the relationship. For example, researchers have aimed to determine which visual features predict where we look next. This is a useful step in determining the neural basis of gaze control (Ramkumar et al. 2016). Traditionally, hand-designed visual features have been used to predict where we look next (Itti and Koch 2001), but recently researchers have more accurately predicted fixation locations using deep learning (Kümmerer, Theis, and Bethge 2014). As another example, researchers have studied how features in the environment predict the songs produced by male Drosophila during courtship (Coen et al. 2016). Using a generalized linear model with a sparsity prior, this study found that the distance between the male and female was the strongest predictor. This allowed the researchers to then investigate the neural pathway that was responsible for distance modulating song amplitude. More accurate behavioral models can allow researchers to better investigate the relationship between neural activity and behavior.

In medicine, it is important to understand the underlying factors that are predictive of disease. This has be done by finding the importance of neuroimaging features in traditional classification techniques (e.g., determining which functional connectivity measures are predictive of Alzheimer's disease in a logistic regression classifier (Challis et al. 2015)). More recently, studies have used a variety of methods to look inside deep learning classifiers to determine the important features (e.g., determining fMRI connectivity relationships that are predictive of ADHD (Deshpande et al. 2015) and schizophrenia (Kim et al. 2016)). This approach has also had success in animal models. For instance, in a mouse model of depression, researchers determined which features (power and coherence) of neural activity in prefrontal cortex and limbic areas were predictive of pathological behavior (Hultman et al. 2016). They were then able to use this information to design a neural stimulation paradigm to restore normal behavior. By using machine learning in each of these applications, the researchers were able to test if their variables predicted disease without having to assume the form of that relationship.

**Neuroanatomy / Structure**

Just as with neural activity, machine learning can help researchers better understand how neuroanatomical features across the brain are predictive of disease. A general approach is to construct an ML classifier to determine whether a subject has the disease, and then look at the importance of the features (e.g., brain areas or connections) in that classifier. In one example, researchers trained an SVM classifier to predict depression based on graph-theory based features derived from diffusion-weighted imaging, and then looked at the importance of those features (Sacchet et al. 2015). In another example, researchers trained a random forest classifier to predict Alzheimer's disease from structural MRI and then determined which brain areas were the most predictive features in this classifier (Lebedev et al. 2014). Another general approach is to compare classification models that are constructed using different features. For example, the previously mentioned paper (Lebedev et al. 2014) also compared classifiers constructed with different feature sets, e.g., one using cortical thickness measures and one using volumetric measures. There are thus multiple ways in which ML can inform us about the predictive relationship between neuroanatomic features and neurological disease.

Neurons have complicated shapes with varying biological structure that vary widely across brain regions and across species (Zeng and Sanes 2017). Many approaches have been proposed to classify neurons: electrophysiology (Teeter et al. 2018), morphology (Vasques et al. 2016), genetics or transcriptomics (Nelson, Sugino, and Hempel 2006), and synaptic connectivity (Jonas and Kording 2015). Machine learning can help with this endeavor (Armañanzas and Ascoli 2015; Vasques et al. 2016). The cell types can be labeled based on one modality (e.g. whether the cell is inhibitory or excitatory), and then these labels can be predicted based on another modality (e.g., morphology). For instance, both (López-Cabrera and Lorenzo-Ginori 2017) and (Mihaljević et al. 2015) have used ML to predict cell type based on morphological features, and investigated the importance of those features. This can both tell us which features are unique across cell types, and also which features are shared (Farhoodi and Kording 2018). In all of these areas, ML can help us to identify important features that shape neurons and transform our view of neuroanatomy.

**Caveats**

When ML methods are used to estimate which variables are predictive, one must be aware of the general caveats of ML methods. These were outlined in Role 1. One should be aware of overfitting, and select the method and regularization technique to maximize the accuracy on held-out data. The higher one's accuracy on a test set, the greater one's statistical power in estimating the predictiveness of a variable.

As mentioned in the introduction to the Role, feature importance methods determine the importance of the raw inputs to the ML model. Finding relevant combinations of features is a separate problem than the one we outline here in Role 2, and often involves looking within the "black box" of ML systems (described in Role 4) or using unsupervised learning methods (Guo et al. 2011; Suk et al. 2015; Längkvist, Karlsson, and Loutfi 2012).

While ML can easily return numbers about predictive relationships, it is important to be careful about the interpretation of these results. Crucially, these methods do not make any claims about causal relationships between variables. If variable X predicts Y, it could be because X causes changes in Y, because Y causally affects X, or because because some variable Z was not observed and affects both X and Y. Additionally, the results about predictive relationships can be highly dependent on the other variables included in the ML model (Stevenson 2018). That being said, these same issues of interpretability exist for any regression method, including simple models.

Determining the importance of features within a ML model is not the only method for determining predictive relationships between variables. For example, information theory has the concept of mutual information. Mutual information also determines how much information one variable has about another variable, and sometimes doesn't require assumptions about the form of the data. The clear meaning of mutual information makes this metric attractive. However, for high-dimensional datasets (like spike trains), without making assumptions about the form of the data, calculating mutual information can be prohibitively time consuming.

## Role 3: Benchmarking simple models

There are many uses of modeling in the biological sciences, and not all uses are satisfied by ML. In particular, many biophysical models embody specific hypotheses about a biological mechanism. The Hodgkin-Huxley model is a canonical model of this type, as the equation itself specifies how ion channel kinematics lead to spikes. Machine learning methods, on the other hand, aim primarily for prediction, and by and large do not automatically build mechanistic hypotheses of how the inputs relate to the outputs. Until methods for model interpretability progress, ML models cannot replace simpler models in this regard.

Simple, hypothesis-driven models are meaningful only to the extent that they are correct. One can easily check a model's accuracy from its predictive performance (e.g. $R^2$), as is often done. However, it is often hard to know how much error derives from sources of noise versus systematic insufficiencies of the model. This is where ML can help. It can serve as an approximate upper bound for how much structure in the data a simpler model should explain. If a human-generated model is much less accurate than ML methods trained on the same task, it is likely that important principles are missing. If, on the other hand, an intuitive model matches the performance of ML, it is more likely (but not guaranteed) that the posited concepts are, indeed, meaningful. Thus, we argue that if a hypothesis-driven model is to be trusted, then it must at least match ML methods trained on the same task.

This approach stands in contrast to the current paradigm of testing a model by comparing it with previous (simple) models. This comparison may be meaningless if both models are very far from the peak ML predictive performance. Even if a new model more accurately explains the data than previous models, it is possible that both models miss important phenomena. Without a change in paradigm, we run the risk of not recognizing predictable complexity when it exists. Benchmarking with ML allows one to check against this trap.

There are great examples of this type of benchmarking from outside of neuroscience. In healthcare, models of patient outcomes must be interpretable enough that caretakers understand what improves outcomes (rather than just predicting the outcome). However, it is also important that the predictions are as accurate as possible. Thus, when researchers made new, interpretable, models of pneumonia risk and hospital readmission, they compared the performance of interpretable models with ML benchmarks (Caruana et al. 2015). In psychology, researchers have compared human-made models against ML benchmarks to understand the limitations of current behavioral models (Kleinberg, Liang, and Mullainathan 2017). This approach should also be advantageous within neuroscience.

Finally, we want to point out that models can be compared against benchmarks on subsets of the data, which can help researchers determine what aspects of their model need to be improved. As an abstract example, imagine that we have a simple model for the activity of a brain area during tasks A and B. The model is close to the ML benchmark for task A, but not B. This tells us that the model needs to be revised to better take task B into account. Thus, using benchmarks can also tell us which components of models need to be improved.

**Neural Activity / Function**

An important part of neuroscience research has been to create simple models of neural activity or how neural activity leads to behavior. A classic example is how V1 receptive fields can be explained by feedforward projections from LGN receptive fields (Hubel and Wiesel 1962). Another classic example is "population vectors" for producing movement (Georgopoulos, Schwartz, and Kettner 1986). That is, if individual motor cortex neurons are treated as vectors towards their preferred movement direction, their vector sum will produce the output movement direction. When these types of models are proposed, it would be beneficial to compare their prediction performance (how well they predict neural activity or behavior) against ML models. This simple supplementary comparison could provide crucial information about how much neural activity or behavior remains to be explained. Unfortunately, benchmarking figures for new models are quite uncommon in neuroscience.

Recent work has demonstrated that ML benchmarks often significantly outperform simple models in neuroscience in terms of predictive performance. Neural networks in particular, have been shown to often describe neural activity far better than traditional, simple models. Neural networks better predict the activity of retinal ganglion cells (McIntosh et al. 2016), primate V4 and IT (Yamins et al. 2014), and auditory cortex (Kell et al. 2018). These results are plain demonstrations of the deficiencies of previous models. An additional benefit of using ML as a benchmark to simple models is that these comparisons can sometimes reveal what is lacking from the current model. For example, if the simple model is a static linear method (as in the population vector example), and the ML method of an RNN provides much better performance, it might be important to update the simple model to include dynamics or nonlinearities. It would be desirable if comparisons to benchmarks were made with the introduction of new simple models.

## Neuroanatomy / Structure

Machine learning can also help benchmark the simple models that describe neuroanatomy. For example, many models have been proposed to describe the complexity of neurons' morphologies. There are simple models describing the relationship between the diameters of segments at branching points (Rall 1964), the linear dependency of branch diameter on its length (Burke, Marks, and Ulfhake 1992), and the fractal dimensions of neurons and their self-similarity (Werner 2010). In all of these examples, it would be possible to use ML techniques on the raw data to create upper performance bounds for these models. This promises to make anatomical modeling more meaningful.

## Caveats

Matching the accuracy of a simple model to an ML model does not by itself guarantee that the simple model captures the explainable structure in the data. The ML model may also miss important aspects. For this reason one should employ the same precautions against ML underperformance that we outlined in Roles 1 and 2. Overfitting, model choice, and dataset size are all important considerations. While ML models are not oracles, it remains the case that a simple model should at least match the performance of ML if it is to be considered complete.

An ML benchmark is not the only way to estimate if a simple model captures the explainable aspects of the data. Sometimes the best benchmark is not a model at all but a biological system. For example, if one is modeling human performance on a visual task, it is more direct to use actual human performance on the task as the benchmark rather than an ML model of vision. In this case, the necessary criteria for the model would be that it reproduces human behavior.

Comparing to an ML benchmark is not the only approach for determining the maximum amount of signal a simple model should explain. One can repeat an experiment several times and observe how much the result varies, and then assume that similarities across experiments are the signal, and the differences are unexplainable noise. In modeling how neural activity changes with stimuli, for example, it is common to present a single stimuli multiple times (Schoppe et al. 2016). The aspects of the neural recording that vary between repetitions are considered unrelated to the stimulus. However, for neural activity, the act of repetition itself can change the response, and this effect must be modeled (Grill-Spector, Henson, and Martin 2006). Furthermore, it is often hard to deliver the same exact stimuli to a moving animal. This drawback is general to this approach of estimating noise: each iteration must be guaranteed to be identical.

Finally, it is important to note that predictive performance is just one of many aspects of models that make them valid. These aspects, like biological plausibility, should of course be taken into account (Kording et al. 2018). High accuracy does not guarantee that a model is a correct representation of the true system. There may be many models that achieve similar performance. For this reason, attaining the maximal predictive performance is a necessary criteria of a model's validity, but certainly not a sufficient one.

## Role 4: Serving as a model for the brain

The role of computational models of the brain is not only to predict, but also to serve as human-understandable distillations of how we think the brain works. It has recently become a more popular idea that deep neural networks are good models of the brain, albeit at a high level of abstraction (Marblestone, Wayne, and Kording 2016; Hassabis et al. 2017; Kietzmann, McClure, and Kriegeskorte 2017). Even a decade ago, this idea seemed less appealing to the field given the hyper-simplified form of contemporary neural networks. However, numerous recent empirical studies have pointed to unexpected parallels between the brain and neural networks trained on behaviorally relevant tasks. Here we review these suggestive studies and discuss the various ways that artificial neural networks are becoming better models of biological ones. While exciting, much work is needed (and is ongoing) to evaluate the extent to which current neural networks are good models of the brain, and on what levels. We discuss neural activity and neuroanatomy without separation, as both aspects are often integrated together in ML models of the brain. In keeping with the theme of this review, we focus on reviewing work centered on supervised learning and omit other learning-based models of the brain like unsupervised learning or reinforcement learning.

The trend of comparing trained neural networks to the brain was reignited recently due to the great achievements of neural networks at behavioral tasks, such as recognizing images (He et al. 2015). Interestingly, these networks have many parallels to the ventral stream in vision. These networks are explicitly hierarchical and multi-layered. Information from image pixels typically is processed through upwards of a dozen layers of "neurons", or nodes. In addition to their analogous organization, their activations are similar. For example, it has been observed that early nodes have Gabor-like receptive fields (Güçlü and van Gerven 2015), reminiscent of the edge detectors seen in V1. Moreover, activations in early/intermediate/later layers of these networks make excellent predictions of V1/V4/IT responses, respectively (of both individual neurons and fMRI responses) (Yamins and DiCarlo 2016; Yamins et al. 2014; Khaligh-Razavi and Kriegeskorte 2014; Güçlü and van Gerven 2015). Recent work has further extended the similarities. Deep neural networks are similarly invariant to viewpoint in object recognition (Saeed Reza Kheradpisheh et al. 2016), respond similarly across images (Khaligh-Razavi and Kriegeskorte 2014), and make similar types of errors (Saeed R. Kheradpisheh et al. 2016). This litany of similarities is longer and extends over a broader range of the visual cortex than any competing class of models.

The similarities between trained neural networks and the brain extend beyond the visual system. The format of these studies, nearly universally, is to compare the internal response properties of a brain area to those of a neural network trained on a behavioral task associated with that brain area. A pioneering study published three decades ago showed the similarity between posterior parietal neurons and a neural network trained to locate objects in a visual scene (Zipser and Andersen 1988). More recently, networks trained on scene recognition could accurately predict responses in the occipital place area (Bonner and Epstein 2018). Networks trained on speech recognition and musical genre prediction have activity similar to the auditory cortex (Kell et al. 2018). Recurrent neural networks trained to reproduce monkey movements contained units with activities very similar in selectivity to neurons in the primary motor cortex (Sussillo et al. 2015). The units of recurrent networks trained on navigation tasks have activations similar to the grid and place cells of the entorhinal cortex and hippocampus (Kanitscheider and Fiete 2017; Cueva and Wei 2018; Banino et al. 2018). The similarity between response properties of artificial neural networks and the brain is a sign that these models may capture important aspects of the brain's computations.

A common drawback of these models is that though they predict activity well, they do not reveal the details about the neural computations being performed. It would be worrisome to replace a brain we cannot meaningfully understand with a neural network equally as complicated. In some cases, it may be possible to tease apart what a neural network has learned and relate it to physiology. Recent work in the retina, for example, has related the activity of a two-layer network fit to ganglion cells to specific upstream cell types (Maheswaranathan et al. 2018). In general, however, the uninterpretability of networks poses a problem, and it has become very attractive to understand the inner workings of neural networks. Recent work includes methods for visualizing what features and information are represented at different scales within convolutional neural networks (Olah et al. 2018) and methods for understanding the dynamics within recurrent neural networks (Sussillo and Barak 2013). In fact, researchers are also developing new model architectures that are more easily interpretable (Foerster et al. 2017; Linderman et al. 2017). Additionally, it is possible to perform neuroscience-inspired experiments on neural networks that would be impossible to perform on a biological brain. For example, researchers recently tested whether the tuning of individual units in neural networks were important for classification generalization (Morcos et al. 2018). Additionally, fully observable neural networks may serve as test-beds for new neural data analysis methods, which are greatly needed to increase understanding (Jonas and Kording 2017). If methods improve to interpret neural networks, they may well challenge assumptions within neuroscience and provide new hypotheses to test.

At the level of learning, brains and neural networks are less apparently similar. It is an open and disputed question whether brains are capable of learning in a similar supervised way as current neural networks. Backpropagation, the method for training neural nets, is not considered by many to be a biologically plausible mechanism for credit assignment (see (Bengio et al. 2015) for reasons it is not biologically plausible). However, this is a subject of debate. One recent paper showed that random feedback weights still allows for successful learning (Lillicrap et al. 2016), solving one of the implausible aspects of backpropagation. Other work has presented networks based on the apical/basal dendrites to solve the problem of credit assignment (Guerguiev, Lillicrap, and Richards 2017; Körding and König 2001; Sacramento et al. 2017). There have been many other recent (Scellier and Bengio 2016; Bengio et al. 2015, 2017) and historic (Hinton and McClelland 1988) works creating more plausible credit assignment mechanisms. However, one challenge is that many biologically-motivated deep learning algorithms do not scale well to large datasets (Bartunov et al. 2018). The eventual resolution to this debate will reveal whether the similarities in response patterns reviewed above emerge from a similar learning rule, or simply a similar solution to problems in cognition.

Another concern about viewing artificial neural networks as models of the brain is that they are not biologically realistic. However, there has been much recent work attempting to create neural networks that have features that are more biologically plausible. One focus on creating biologically plausible neural networks is on having spiking (binary), as opposed to continuous, units. Many recent research papers have begun to create spiking neural networks that successfully solve typical machine learning (ML) problems (Zenke and Ganguli 2017; Nicola and Clopath 2017; Mozafari et al. 2018; Bellec et al. 2018). There has also been recent work developing architectures that are more biologically realistic (Costa et al. 2017; Linsley et al. 2018), for example those inspired by cortical microcircuits (Costa et al. 2017). Work on biologically-plausible deep learning will help to address how much artificial networks should be seen as faithful models of the brain.

The analogy between brains and networks invites a change in research focus for neuroscience. One might consider, for instance, focusing on which cost functions the brain is optimizing rather than the final properties of the network. Similarly, it is important to focus on determining the learning rules implemented by the brain. We have discussed some of these matters recently in a separate review, particularly focusing on how neuroscience should learn from machine learning's quest to understand neural networks (Marblestone, Wayne, and Kording 2016).

**Caveats**

Neural networks are models of the brain only at a certain level of abstraction. What this level is, and how much of brain function this includes, is a matter of critical debate. Certainly, as Lake et al. argue, brains can do things that backpropagation-trained neural networks appear incapable of (Lake et al. 2017). Neural networks require large amounts of data to train, for example, while the brain can often learn from few examples (Carey and Bartlett 1978; F. Xu and Tenenbaum 2007). The existence of "adversarial examples" that are misclassified by neural networks but not humans are another common example of outstanding functional differences (Nguyen, Yosinski, and Clune 2015). Thus, even at the level of function there are outstanding differences.

The parallels between response properties in neural networks and the brain were surprising in part because they indicated that the two systems were not only both classifying objects correctly, but were doing so in a similar way. That is, they had similar implementations of a given function. At the level of implementation, however, neural networks and brains are known to be dissimilar in many ways. Neural networks have no analogs for the array of unique cell types, neuromodulators, synapse-specific adaptation and short-term plasticity, or precise spike timing. It is important to recall as well that several neural network architectures can all predict activity reasonably well, despite being different in form (Yamins et al. 2014). These various differences suggest that we can only take deep networks to be good models of the brain at a high, functional level.

A consensus opinion is that much learning in the brain does not occur via supervised ML. Much of what the brain does may be closer to unsupervised learning (Hinton and Sejnowski 1999; Doya 2000) or reinforcement learning (Wang et al. 2018; Gläscher et al. 2010; Glimcher 2011; Doya 2000). This again makes it somewhat surprising that, in many areas, pretrained NNs are the best models yet of activity. Thus, the similar response properties of neurons and networks may often arise through different processes. The early layers of trained convolutional neural network, for example, have orientation tuning like that seen in V1. However, orientation tuning in V1 is also consistent with bottom-up self-organization (an unsupervised paradigm) (von der Malsburg 1973; Olshausen and Field 1997; Bell and Sejnowski 1997). For now, it is most important to allow the possibility that similar results can derive from different processes.

These differences make it difficult to be precise about at what level of abstraction neural networks are good models of the brain. In each domain of learning, physiology, and anatomy, the answer is slightly different. A central question fueling the next decade of research, then, will be determine which details of biology neural networks effectively abstract out, and which details add a fundamentally different sort of function than neural networks currently include.

## Box 2: Example of Supervised ML in Practice

**The data.** To train a supervised ML method, one needs to have input data and labeled outputs. On the right, we show EEG traces (Brinkmann et al. 2016). Let's say the goal is to detect when a seizure is occurring. The data has been labeled to show the output: blue for a normal brain state and red for a seizure.

**Split into training/testing tests.** It is important to determine the accuracy of an ML model on separate data than was used to train the model. This is because a good model should generalize to new data and not just memorize training examples. A common technique is cross-validation, which repeatedly splits the data into different training/testing sets.

**Preprocess / extract features from the data.** Often, it is helpful to extract features from the raw data to use as inputs to the ML algorithm (e.g., for EEG, using power within multiple frequency bands rather than the full energy spectrum). Many modern ML techniques are making feature extraction less crucial. For example, raw images are often fed into convolutional neural nets. The same data preprocessing steps need to be performed on the training/testing sets.

**Model selection.** There are many different ML algorithms, so which should we use? First, it depends on the data type. For instance, convolutional neural nets are good for images and recurrent neural nets are good for sequential data. It also depends on the amount of data. With limited data, models with fewer parameters are often helpful, although complex models can still be successful using techniques such as regularization. In general, we have found that tree-based methods (Random Forest and XGBoost) and feedforward neural nets with a couple hidden layers work well with default parameters for many problems. Finally, it is possible to not choose and use many algorithms (see ensembling below).

**Hyperparameter optimization.** Beyond the parameters that models fit, many ML models have hyperparameters, which generally relate to the model structure or how the model is fit. For example, neural networks can have different numbers of units in the hidden layer. The general strategy is to further split the training set into training and validation sets, and determine the hyperparameters that lead to the best fit on the validation set. Toolboxes exist that will intelligently find good hyperparameters (e.g., Hyperopt and BayesianOptimization).

**Fit the model.** We fit the model and make predictions on the test set. This model can be used for any of the 4 roles!

**Ensembling.** Ensemble methods are a common approach for maximizing performance that are often used in ML competitions, such as Kaggle. Ensembles combine predictions from multiple different ML models (see Box 1).

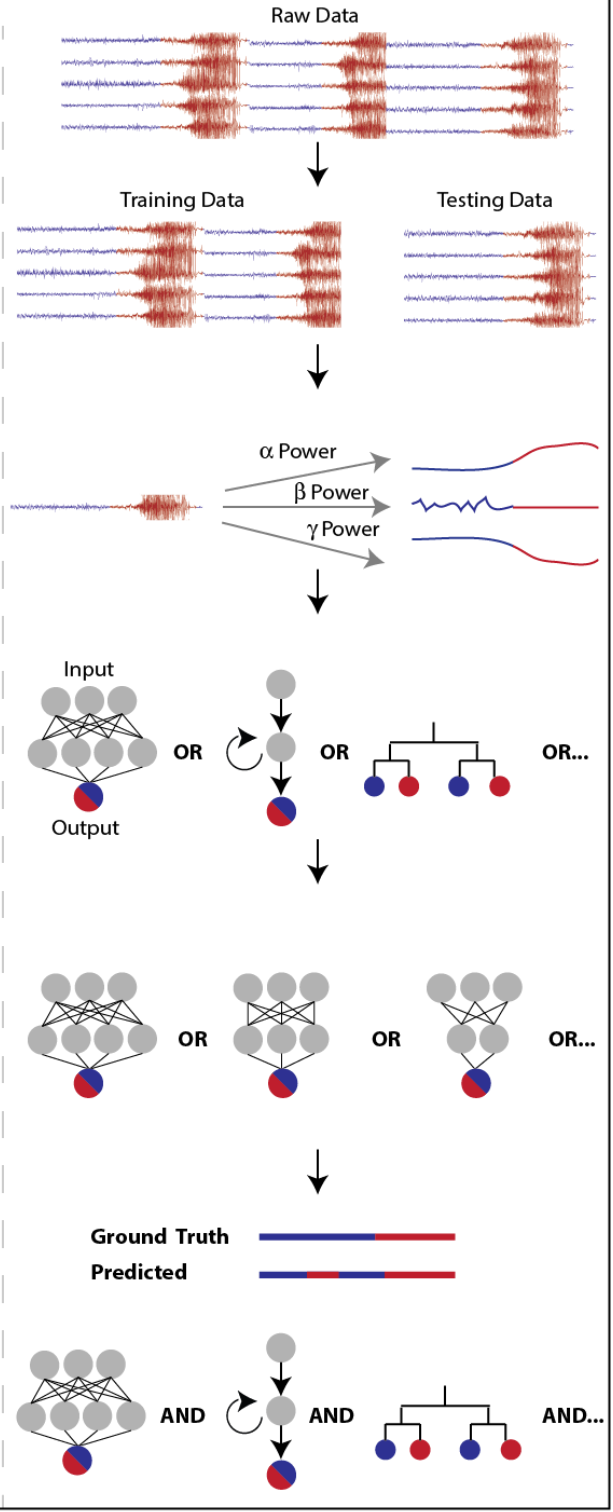

## Discussion:

Here we have argued that supervised machine learning (ML) has four primary roles within system neuroscience: 1) Solving engineering problems; 2) Identifying predictive variables; 3) Benchmarking simple models; and 4) Serving as a model of the brain. As the current trend in applying ML to neuroscience (Fig. 1) indicates, we believe that the influence of ML on neuroscience will continue to grow.

As is the case with any modeling, we wish to remind readers that it is important to be cautious when interpreting ML models. High predictability does not mean causality (Pearl 2009; Katz et al. 2016). This is especially true because there are so many unobserved variables within neuroscience. For instance, the apparent importance of a variable within a model may be inaccurate due to other unobserved variables (Stevenson 2018). Moreover, high model accuracy does not mean that it is actually a causally correct model of what is occurring in the brain. High accuracy is a necessary, but not a sufficient condition for model correctness. This is because there is an enormous space of potential models that could explain the data well. This is a difficulty of modeling the brain and identifying predictive variables with ML, but does not impact the use of ML for engineering or benchmark applications.

Machine learning extends beyond supervised learning and thus plays a greater role in neuroscience than we describe here. Using unsupervised learning methods, one can reduce the dimensionality of data (Cunningham and Yu 2014; Gao and Ganguli 2015), use clustering algorithms to discover new classes/categories (Armañanzas and Ascoli 2015; Drysdale et al. 2017), create generative models (Molano-Mazon et al. 2018; Arakaki, Barello, and Ahmadian 2017), and extract features automatically (Guo et al. 2011; Suk et al. 2015; Längkvist, Karlsson, and Loutfi 2012). Moreover, reinforcement learning (RL) is another ML category used within neuroscience. RL is inspired by behavioral psychology and can be used as a model of the brain to understand the mechanisms of reward-based learning and decision making (Wang et al. 2018; Gläscher et al. 2010; Glimcher 2011; Doya 2000). The four roles of this paper are all about supervised learning and do not include unsupervised learning and RL.

As neurotechnologies advance, the role of ML in neuroscience is likely to continue to grow. Datasets are rapidly growing (Stevenson and Kording 2011; Kandel et al. 2013) and becoming more complex (Glaser and Kording 2016; Vanwalleghem, Ahrens, and Scott 2018). Machine learning will be needed for this regime of data, as after all, there is only so much time a human being can spend looking at data. Moreover, as datasets get bigger, ML techniques become more accurate. Additionally, it is hard to reason about complex and high-dimensional datasets. Of all of the models that could explain a complex system, it is possible to think only about those models that are simple enough to imagine – to outline in human working memory. But in biology, as opposed to physics, there are good reasons to assume that truly meaningful models must be fairly complex (O'Leary, Sutton, and Marder 2015). While humans will correctly see some structure in the data, they will miss much of the actual structure. It is simply difficult to intuit models of nonlinear and recurrent biological systems. In these situations, it may be necessary to seek help from ML methods that can extract meaningful relationships from large datasets.

The use of ML will also continue to expand as ML gets easier to use. Applying ML has already become fairly straightforward. At application time, one requires a matrix of training features and a vector of known labels. Given the availability of the right software packages (Pedregosa et al. 2011), generally, only a few lines of code are then needed to train any ML system. In fact, there has been recent work on automated ML (Feurer et al. 2015; Kotthoff et al. 2017; Guyon et al. 2015; Elsken, Metzen, and Hutter 2018), so that users do not need to make any decisions on specific methods to use, how to preprocess the data, or how to optimize hyperparameters. Thus, it is becoming less important for neuroscientists to know the details of the individual methods, which frees them to focus on the scientific questions that ML can answer.

The power of ML allows for the design of new types of experiments. Experiments will benefit from obtaining as much data as possible, measured both by the number of samples and number of dimensions. Since ML promotes experimental approaches that aim for predictions rather than for interpretations of the role of each variable, one can record as many variables as will improve predictions. This stands in contrast with the traditional scientific method of variable interpretation as, through multiple comparison testing corrections, it is not possible to say much about any variable if too many are recorded. Machine learning can also be used to optimize experimental design, e.g. by intelligently choosing stimuli that will maximize firing rates (Cowley et al. 2017) (here acting in Role 1 as an engineering tool). Researchers should keep ML methods in mind when designing experiments.

For a long time, neuroscientists have worked on improving ML techniques, and many ML techniques have been inspired by thoughts about brains and neural computation. With the growth of the ML field, the flow of information is becoming multi-directional. While neuroscience continues to inspire ML development, ML is also on the way to becoming one of the central tools and concepts in neuroscience.